\documentclass[aps,twocolumn,showpacs,eqsecnum]{revtex4}

\usepackage{natbib}
\usepackage{dcolumn}
\usepackage{amsmath}
\usepackage{graphics}
\usepackage{color}

\begin{document}

\title{Raman process under condition of radiation trapping\\ in a disordered atomic medium}

\author{L.V. Gerasimov, V.M. Ezhova, D.V. Kupriyanov}%
\affiliation{Department of Theoretical Physics @ St.-Petersburg State Polytechnic University\\ 195251, St.-Petersburg, Russia}%

\email{kupr@dk11578.spb.edu}%

\author{Q. Baudouin, W. Guerin, R. Kaiser}%
\affiliation{Institut Non Lin\'{e}aire de Nice, CNRS, Universit\'{e} de Nice Sophia-Antipolis, 1361 route des Lucioles, 06560
Valbonne, France}%

\date{\today }

\begin{abstract}
 We consider the Raman process developing in a disordered medium of alkali-metal atoms when the scattered modes are trapped on a closed transition. Our theoretical analysis, based on numerical simulations of the Bethe-Salpeter equation for the light correlation function, which includes all Zeeman states and light polarization, lets us track the stimulated amplification as well as the losses associated with the inverse anti-Stokes scattering channel. We discuss possible conditions when this process could approach the instability point and enter the regime of random lasing.
\end{abstract}


\maketitle%


\section{Introduction}

\noindent Light transport in an ultracold and optically thick atomic gas has been intensively explored in experiment and theory over the past decade. This has stimulated a vast amount of innovative research in a wide range of areas including quantum optics, precision measurements and quantum information science \cite{FIM,PSH,Simon}. The Raman-type interaction in $\Lambda$-configured atomic systems forms the basis of many schemes of quantum interface between light and atoms. Remarkable progress in developing one dimensional protocols of Raman-based quantum memories \cite{NGPSLW,HSCLB,CDLK,MSLOFSBKLG}, entanglement creation and deterministic quantum interface \cite{KMJWPCP} motivates further extensions towards more general three dimensional configurations. The problem of light-matter quantum interface between spatial modes of light and matter is intensively discussed in the context of multimode quantum memory scalability and quantum imaging \cite{SorSor,VSP,VNGMSGL}. As has been recently pointed out  in \cite{GSKH} the light propagation through an atomic sample under multiple elastic scattering process could be coherently controlled with an external mode and implemented as a mechanism of atomic memories, which potentially would be more effective than the standard protocols assuming a one-dimensional propagation channel.

In the present report we consider one specific experimental configuration where the combination of microwave and optical fields would initiate the Raman process in a disordered and optically thick atomic gas. If this process evolves in the hyperfine manifold of the D$_2$ line of an alkali-metal atom then the Raman-scattered mode can match the frequency of a closed dipole-transition, such that the scattered light becomes trapped and shielded from its direct propagation, a phenomenon called ``radiation trapping'' \cite{Molisch}, already studied in cold atoms \cite{Fioretti:1998,Labeyrie:2003}. In such a scenario we can follow the conversion of the hyperfine repopulation into the trapped light mode and investigate the quantum interface between the spin and optical modes. The primary goal is to follow how the light trapping modifies the Raman process.  As we show, in some conditions, the system can demonstrate the stimulated Raman amplification of the emitted light, but the amplification process competes with the losses via inverse anti-Stokes scattering channel. Our calculation tracks the balance in light redistribution between both channels.

Besides quantum-optics applications, the model presented in this paper, which notably includes the atomic Zeeman structure and light polarization, may also be useful to describe experiments on random lasing in cold atoms. A random laser is a laser without cavity, in which the feedback effect is provided by multiple scattering (i.e., radiation trapping) in the gain medium itself \cite{Letokhov,Cao,Wiersma}. Raman gain has been used to sustain lasing based on cold atoms as the gain medium in several experiments \cite{Hilico:1992,McKeever:2003,Guerin:2008,Vrijsen:2011,Bohnet:2012} and the combination of Raman gain and multiple scattering, leading to the random laser regime, has recently been demonstrated in \cite{BMGGK}, using a scheme not very different from the scheme that we consider here. Both schemes share the property that Raman gain occurs at a frequency that is in resonance with another atomic line which provides multiple scattering. However, in \cite{BMGGK}, Raman gain is based on a population inversion obtained by a strong incoherent optical pumping, whereas here we consider a weak coherent microwave pumping, which does not provide a population inversion, and Raman gain is not strong enough to compensate losses due to inelastic scattering. We shortly discuss, at the end of the paper (Sec. \ref{Sec3C}), some possibilities to overcome this limitation.


\section{Theory}

\subsection{Physical scheme and basic assumptions}\label{II.A}

\begin{figure}[tp]
{$\scalebox{0.6}{\includegraphics*{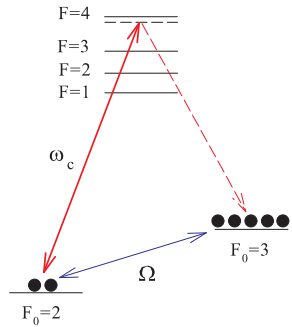}}$ }
\caption{(Color online) Energy levels and the excitation diagram for the Raman process initiated in an ensemble of ${}^{85}$Rb atoms via $F_0= 2 \rightarrow F=2,3\rightarrow F_0=3$ transitions. The Raman emission is a result of the simultameous action of microwave $\Omega$ and optical $\omega_c$ excitations, both are linearly polarized along the quantization direction. The emitted light is trapped on the closed $F_0=3\to F=4$ transition in the optically thick atomic ensemble. While propagating through the atomic sample this light stimulates additional Raman emission and may increase the output fluorescence emerging the sample.}
\label{fig1}%
\end{figure}

\begin{figure}[tp]
{$\scalebox{0.8}{\includegraphics*{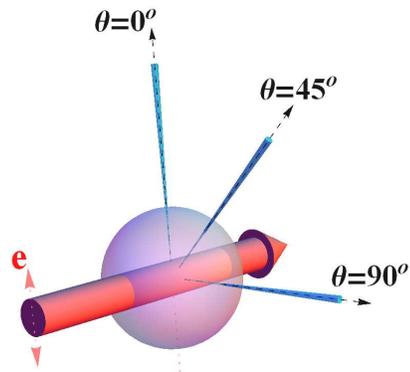}}$ }
\caption{(Color online) Excitation geometry: the linearly polarized control mode (shown as a pink arrow-tube) and microwave field (not shown) initiate the Raman emission for the process shown in figure \ref{fig1}. The observation channels are parameterized by the polar angle $\theta=0^0,\,45^0,\,90^0$ from the control mode polarization direction.}
\label{fig2}%
\end{figure}

The energy diagram and the excitation scheme considered in this paper are based on $^{85}$Rb, as specified in Figs. \ref{fig1} and \ref{fig2}. The rubidium atoms initially populate the upper hyperfine sublevel $F_0=3$ of their ground state. Then a part of the atoms is repopulated to the lower hyperfine sublevel $F_0=2$ with the aid of a microwave field applied between these sublevels. In addition a strong coherent field (control mode) tuned to the forbidden $F_0=2\to F=4$ transition can generate the off resonant Raman process via $F=3,2$ upper states with transporting atoms back to sublevel $F_0=3$. As in \cite{BMGGK}, the crucial feature of such an excitation scheme is that the emitted light will be immediately trapped by slow diffusion (multiple scattering) on the closed $F_0=3\to F=4$ transition if the atomic sample is optically thick \cite{Labeyrie:2003}. Then this portion of light, diffusely trapped in the disordered atomic ensemble, can further stimulate additional Raman emission. However such Raman amplification is strongly reduced since a significant part of the trapped light will be lost via the inverse anti-Stokes scattering channel as a result of its interaction with the Autler-Townes resonance structure created by the control mode. The light dynamics will track the balance between the Raman-gain mechanism and the losses accompanying its diffusion transport. 

Let us specify the basic assumptions, which we will follow in our consideration. The dipole-type interaction $\hat{V}(t)$ of atoms with the optical mode initiating Raman process will be taken in a linear polarization along the $Z$-axis of the laboratory frame. The interaction $\hat{U}(t)$ of the magnetic dipole moment with the microwave field will be considered as linearly polarized along the same direction. If in their order of magnitude the matrix elements of interactions with optical and microwave modes are roughly estimated as $\bar{V}$ and $\bar{U}$ respectively, then in our discussion we will mostly constrain them by the following inequality: $\gamma\gg \bar{V}^2\gamma/\Delta_{\mathrm{hpf}}^2\gg \bar{U}$, where $\Delta_{\mathrm{hpf}}$ is the hyperfine splitting in the upper state and $\gamma$ its natural linewidth. This inequality justifies that the rate of spontaneous decay is greater than the rate of spontaneous Raman scattering, but the latter is greater than interaction with the microwave field. In other words the interaction with the microwave field is the weakest process in the system and only a small number of atoms can be repopulated into sublevel $F_0=2$ from the sublevel $F_0=3$. It thus prevents from reaching a high Raman gain based on population inversion as in \cite{BMGGK}.

With these assumptions the expansion of the emitted light in multiple scattering medium results in that any point-like atomic scatterer transforms the locally propagating plane mode into a spherical wave. The energy, polarization and coherent properties of the radiation, originally emitted in the Raman process, can be entirely reproduced via transformation of its correlation function inside the medium when light propagates and emerges the sample,
\begin{equation}
D^{(E)}_{\mu\mu'}(\mathbf{r},t;\mathbf{r}',t')=\left\langle \hat{E}_{\mu'}^{(-)}(\mathbf{r}',t')\hat{E}_{\mu}^{(+)}(\mathbf{r},t)\right\rangle ,
\label{2.1}
\end{equation}
where $\hat{E}_{\mu}^{(\pm)}(\mathbf{r},t)$ are the positive/negative frequency components of the Heisenberg operators of the electric field associated with this light.

If the light is locally a stationary plane wave freely propagating in direction $\mathbf{n}$, then the corresponding spectral energy distribution can be expressed by Fourier transforming the above correlation function over $\tau=t-t'$,
\begin{eqnarray}
I(\mathbf{n},\mathbf{r},\omega)&=&\frac{c}{2\pi}\sum_{\mu=x,y}%
\int_{-\infty}^{\infty}d\tau\,\mathrm{e}^{i\omega\tau}%
\nonumber\\%
&&\times D_{\mu\mu}^{(E)}(\mathbf{r},\bar{t}+\tau/2;\mathbf{r},\bar{t}-\tau/2) ,%
\label{2.2}%
\end{eqnarray}
where $\bar{t}$ is considered here as a time argument dependence, which vanishes in steady-state conditions. For time-dependent process it is more natural to introduce the Poynting vector associated with the light energy transported in direction $\mathbf{n}$. This can be done via tracing the correlation function over its polarization components $\mu=x,y$ in the plane orthogonal to $\mathbf{n}$ and considering it in coincident spatial points and time moments,
\begin{equation}
I(\mathbf{n},\mathbf{r},t)=\frac{c}{2\pi}\sum_{\mu=x,y}%
D_{\mu\mu}^{(E)}(\mathbf{r},t;\mathbf{r},t).%
\label{2.3}%
\end{equation}
These two alternative representations can be compromised to the Wigner-type representation, which considers the instantaneous part of the field energy flux for a certain part of its spectrum as distributed near the point $\mathbf{r}$ and propagating in direction $\mathbf{n}$. The Wigner function would be given by the same Fourier expansion as (\ref{2.2}) but implies the non-vanishing dependence on $\bar{t}$.

For dilute atomic system when $n_0\lambdabar^3\ll 1$, where $n_0$ is the density of atoms and $\lambdabar=\lambda/2\pi$ is the inverse wavevector of the radiation, the near field effects in the interatomic interaction can be safely ignored. Then, as in the vacuum case, the transverse electric field is given by the time derivative of the vector potential and in our consideration we accept the following approximation
\begin{eqnarray}
\hat{\mathbf{E}}^{(0,+)}(\mathbf{r},t)&=&-\frac{1}{c}\frac{\partial}{\partial t} \hat{\mathbf{A}}^{(0,+)}(\mathbf{r},t)\to%
+\frac{i\bar{\omega}}{c}\hat{\mathbf{A}}^{(0,+)}(\mathbf{r},t),%
\nonumber\\%
\hat{\mathbf{E}}^{(0,-)}(\mathbf{r},t)&=&-\frac{1}{c}\frac{\partial}{\partial t} \hat{\mathbf{A}}^{(0,-)}(\mathbf{r},t)\to%
-\frac{i\bar{\omega}}{c}\hat{\mathbf{A}}^{(0,-)}(\mathbf{r},t),%
\nonumber\\%
\label{2.4}%
\end{eqnarray}
which we introduce for the field operators in the interaction representation and associate $\bar{\omega}$ with a typical carrier frequency identical for all the optical modes. With this simplification we can link the correlation function with the standard Green's function of the quantum electrodynamics approach \cite{BrLfPt}.

\subsection{Diagram analysis and calculation scheme}\label{II.B}

Referring to the diagram of figure \ref{fig1}, the probe photons, emitted by the Raman process near the $F_0=3\to F=4$ ``closed'' transition, is expected to further propagate through the sample in a diffusion regime. In dilute configuration the smoothed dynamics of the Heisenberg operators of the quantized field can be approached by a sequence of independent local transformations on each scatterer forming a bulk medium \cite{SorSor,KSSH}. After mesoscopic averaging of the perturbation theory expansion for the light correlation function it can be rearranged in a series of partial contributions interpreted as multiple scattering paths of certain orders. Then the light transport can be visualized as a combination of such zigzag-type propagation paths, where each segment of the plane dynamics obeys the macroscopic Maxwell description and the propagation is depleted by incoherent scattering creating the secondary, tertiary and higher order scattering waves.

Neglecting the interference among any contributing scattering paths, the entire process can be approximated by a ladder structured Bethe-Salpeter-type expansion for the complete correlation function, which has the following graph description
\begin{eqnarray}
\lefteqn{\left\langle \hat{A}_{\mu'}^{(-)}(\mathbf{r}',t')\hat{A}_{\mu}^{(+)}(\mathbf{r},t)\right\rangle}%
\nonumber\\%
&&=\left\langle\tilde{T}\!\left(\!S^{\dagger}\hat{A}_{\mu'}^{(0,-)}(\mathbf{r}',t')\!\right)%
T\!\left(\!S\hat{A}_{\mu}^{(0,+)}(\mathbf{r},t)\!\right)\right\rangle\Rightarrow%
\nonumber\\%
&&{\scalebox{0.9}{\includegraphics*{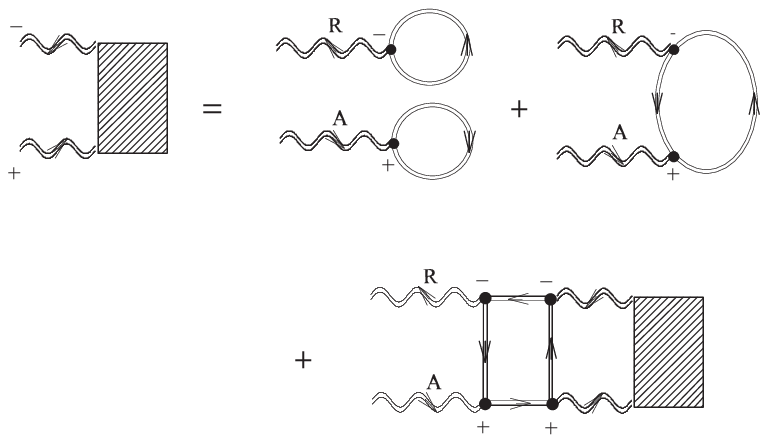}}}%
\nonumber\\%
\label{2.5}
\end{eqnarray}
Here in the second line we introduced the evolutionary operator $S=S(\infty,-\infty)$ expressed by the standard electric-type and magnetic-type dipole interactions between atomic and field subsystems. In the rotating wave approximation this operator generates the set of diagrams, which, neglecting all the interference terms, can be regrouped in the above ladder-type closed diagram equation. In the diagrams each vertex has either positive or negative sign, depending on the interaction terms from either expansion of $S^\dagger$ (pointed by ``+'') or $S$ (pointed by ``-''). The diagram technique, we follow in this paper, was originally proposed for non-equilibrium systems in \cite{Keldysh} and its general formalism is consistently introduced in textbooks \cite{BrLfPt}, for its application to light interaction with atomic systems see also \cite{BatKS}. The Bethe-Salpeter equation, considered in analytical form, is quite cumbersome but its blocking elements have clear physics and are interpreted below.

The outward doubly wavy lines in this equation respectively introduce the exact retarded-type $(R)$ and advanced-type $(A)$ photon propagators in the nonlinear optical medium.
The retarded-type propagator (more rigorously its positive frequency component) is given by
\begin{eqnarray}
\lefteqn{i{\cal D}_{\mu_1\mu_2;+}^{(R)}(\mathbf{r}_1,t_1;\mathbf{r}_2,t_2)=}%
\nonumber\\%
&&\left\langle\left[\hat{A}_{\mu_1}^{(+)}(\mathbf{r}_1,t_1),\hat{A}_{\mu_2}^{(-)}(\mathbf{r}_2,t_2)\right]\right\rangle%
\theta(t_1-t_2)%
\label{2.6}%
\end{eqnarray}
and the advanced propagator can be defined as ${\cal D}_{12;+}^{(A)}={\cal D}_{21;+}^{(R)*}$. These propagators reproduce the mesoscopically averaged plane wave dynamics of any fragment of the probe light controlled by external driving optical and microwave modes.

As shown in appendix \ref{Appndx_A} the photon propagator can be found in the closed form and expressed via the local dielectric susceptibility of the medium. The susceptibility tensor depends on external fields and parameters such as the transition dipole moments $\left(d_{\mu}\right)_{nm}$ (where $\mu=x,y,z$), the local density of atoms $n_0(\mathbf{r})$ at point $\mathbf{r}$, and on the steady state density matrix components $\bar{\rho}_{mm'}$ and $\bar{\rho}_{nn'}$ of the lower and upper atomic states respectively. The atomic repopulation driven by the control and microwave modes provides the stimulated Raman amplification along plane wave propagation between scattering events.

The critical element contributing to the kernel block of the integral equation (\ref{2.5}) is the scattering tensor or amplitude, which has the following graph definition
\begin{equation}
\raisebox{-0.5 cm}{\scalebox{0.8}{\includegraphics*{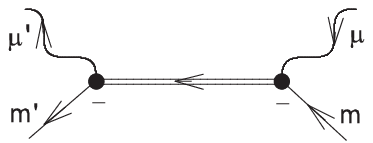}}}%
\sim \alpha_{\mu'\mu}^{(m'm)}(\omega)|m'\rangle\langle m|%
\label{2.7}%
\end{equation}
and is given in its analytical form by
\begin{eqnarray}
\alpha_{\mu'\mu}^{(m'm)}(\omega)|m'\rangle\langle m|\!&=&\!%
-\frac{1}{\hbar}\sum_{n,n'}\left(d_{\mu'}\right)_{m'n'}%
\left(d_{\mu}\right)_{nm}\,|m'\rangle\langle m|%
\nonumber\\%
&&\times {\cal G}_{n'n}^{(R)}(\hbar\omega+E_m+i0)%
\label{2.8}%
\end{eqnarray}
This tensor determines the amplitude of incoherent scattering of the probe photon of frequency $\omega$ in arbitrary direction via either elastic or inelastic channels associated with atomic transition $|m\rangle\to |m'\rangle$, and $E_m$ is the energy of the atomic input state. The presence of the retarded type atomic propagator ${\cal G}_{n'n}^{(R)}(E)$ indicates that the internal spectrum of atomic scatterer is strongly modified by the presence of the external driving optical and microwave modes, see appendix \ref{Appndx_A}.

The product of the scattering tensor with its Hermitian conjugated counterpart forms the diagram kernel block expressing the transformation of the light correlation function in a single scattering event. However only for the photons scattered elastically, when both the $|m\rangle$ and $|m'\rangle$ states belong to the same hyperfine sublevel $F=3$, the transformation can be further converted into the ladder-type diagram sequence and build the closed graph equation. Once a probe photon is scattered inelastically via inverse anti-Stokes channel to the mode with frequency near the forbidden transition $F_0=2\to F=4$  it does not undergo multiple scattering and escapes the medium. Thus the last term in graph equation (\ref{2.5}) is responsible for successive multiple re-scattering of the original light in the nonlinear and optically dense medium.

Because of the pumping process both the input and output states in the elastic channel are actually unstable and the above concept gives us only the first approximation in the calculation of the kernel block in the integral equation (\ref{2.5}). Correction could be done via relevant ``dressing'' of the inward and outward atomic lines in the diagram (\ref{2.7}) by the interaction with the microwave and control modes, which reproduces the population dynamics on the long term scale. In the graph form of the Bethe-Salpeter equation this dressing is indicated by doubling of the atomic lines linking the upper (retarded type) and lower (advanced type) branches of the diagram equation. But with the considered assumptions, where depletion of the $F=3$ is weak and population dynamics is slow, such a correction would be mostly negligible and will be further ignored.

The source terms in equation (\ref{2.5}) consist of two contributions. The first term, which has factorized form, performs the coherent generation of the Raman mode in the three-wave mixing process: $\omega=\omega_c-\Omega$. Since the input and output optical modes are associated with electric-type dipole transitions (polar vectors) but the microwave mode with magnetic-type dipole transition (axial vector), and all three waves have the same polarization, this process is forbidden if the atoms symmetrically populate the Zeeman states of different signs. Such excitation channel would be allowed if the atomic medium had a spin angular momentum polarization in its ground state. The second term corresponds to the process of spontaneous Raman scattering on the atoms populating the sublevel $F_0=2$ as shown in figure \ref{fig1} and provides the actual contributing fluorescence source.

Our calculation scheme consists of the following steps. First we calculate the basic diagram blocks contributing the above graph equation for the correlation function such as the photon propagators, the dielectric susceptibility of the medium, the scattering tensor, etc. Then we make Monte-Carlo numerical simulations of the equation for a sample of given optical depth and atomic distribution. The details of these simulations are further clarified in section \ref{Sec.III.B} and appendix \ref{Appndx_B}. The steady state components of the atomic density matrix driven by external fields can be found either analytically or in general case numerically and are assumed to be given by single atom physics, dressed by the external control field and microwave radiation, mostly neglecting any backaction from the emitted and trapped light. But we assume in our consideration that the trapped radiation would equalize the population distribution over the Zeeman sublevels of the hyperfine state $F_0=3$ via elastic rescattering process.

\section{Results}

\subsection{The susceptibility spectra, extinction and scattering lengths}

In the macroscopic Maxwell theory the electromagnetic properties of the medium are described by its dielectric susceptibility responding on a forwardly propagating probe plane wave. The ``dressing'' processes associated with the external driving optical and microwave fields modify significantly the original structure of the susceptibility and make the medium anisotropic. The closed expression for the tensor of the dielectric susceptibility is derived in appendix \ref{Appndx_A} and given by Eq. (\ref{A.8}). If the driving fields are both linearly polarized then the tensor has the following diagonal form in its main frame representation,
\begin{equation}
 \hat{\chi}=\left(\begin{array}{ccc}
 \chi_{\bot}&0&0\\
 0&\chi_{\bot}&0\\
 0&0&\chi_{\parallel}
 \end{array}\right),
\label{3.1}
 \end{equation}
where the $\chi_{\parallel}=\chi_{\parallel}(\mathbf{r},\omega)$ component determines the response of the atomic polarization on its excitation by the probe mode linearly polarized in the direction of the $Z$-axis (along the external fields polarization) and the transverse component $\chi_{\bot}=\chi_{\bot}(\mathbf{r},\omega)$ gives the response of the probe for any polarization in the orthogonal $X,Y$-plane.

\begin{figure}[tp]
{$\scalebox{0.9}{\includegraphics*{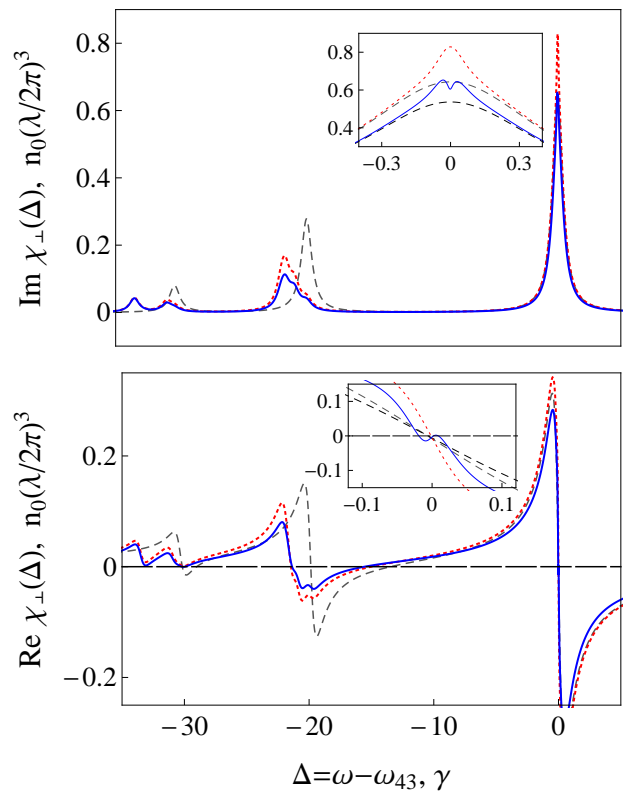}}$ }
\caption{(Color online) Spectral dependence of the imaginary part (upper panel) and real part (lower panel) of the $\chi_{\bot}(\omega)$ component of the susceptibility tensor in units of the dimensionless local density of atoms $n_0\lambdabar^3$. The gray dashed curve reproduces the original undisturbed multiplet of the D$_2$ line of $^{85}$Rb probed from the $F_0=3$ ground state hyperfine sublevel [Fig. \ref{fig1}]. The red dashed curve indicates how this multiplet is modified by the interaction with the optical control mode and the blue solid line shows the spectral profile affected by both the optical and microwave modes. The Rabi frequency for the optical transition has a value of $2\bar{V}= 25\gamma$, for the microwave mode it was selected for the ``clock'' transition $F_0=2,M_0=0\to F_0=3,M_0=0$ and chosen as $2\bar{U}= 0.05\gamma$. The details of spectral behavior near the $F_0=3\to F=4$ resonance are magnified in insets, where the black dashed curve shows how the isotropic part (gray dashed) is reduced by depopulation of the $F_0=3$ state.}
\label{fig3}%
\end{figure}%

\begin{figure}[tp]
{$\scalebox{0.9}{\includegraphics*{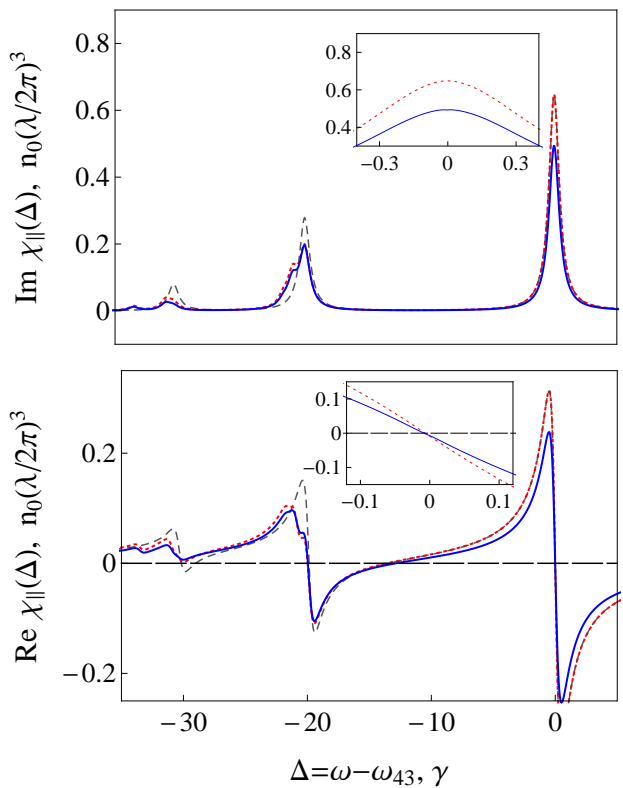}}$ }
\caption{(Color online) Same as in Fig. \ref{fig3} but for the $\chi_{\parallel}(\omega)$ component  of the susceptibility tensor. This component is insensitive to the external fields near the $F_0=3\to F=4$ resonance, where it only tracks the depopulation of the $F_0=3$ state. Other resonances, associated with $F_0=3\to F=3,2$ transitions, are split by the control mode, which partly resolves their Zeeman structure.}
\label{fig4}%
\end{figure}

The typical spectral behaviors of both the major components of the sample susceptibility are shown in Figs .\ref{fig3} and \ref{fig4}. The plotted graphs demonstrate how the original Lorentzian-shaped absorption and dispersion profiles of the D$_2$ line of $^{85}$Rb are modified by the driving optical and microwave modes. In the case of the $\chi_{\bot}$ component the switching-on of only the optical mode creates an additional resonance feature, which overlaps with the $F_0=3\to F=4$ absorption line. In the presented calculations the Rabi frequency of the control light was chosen as $2\bar{V}\simeq 25\gamma$, where we estimated the transition dipole moment by its reduced matrix element for the D$_2$-line. For such a strong field the ground state $F_0=2$ has a light shift and to match resonances the frequency of the control mode has to be slightly red detuned from $\omega_{42}$. Other absorption lines for $F_0=3\to F=3,2$ transitions are also highly affected by such a strong optical mode. The field induces different light shifts for different Zeeman sublevels and splits the lines in several resonances.  We observe here the Autler-Townes resonance structure for the whole upper hyperfine manifold of ${}^{85}$Rb.

After switching on the microwave mode the atoms partly repopulate the second hyperfine sublevel of their ground state and the susceptibility spectrum becomes modified by the external fields ``dressing'' interactions as well as by repopulation dynamics. In our calculations the microwave field was tuned on the $F_0=2\to F_0=3$ hyperfine transition taking into account the light shift. The Rabi frequency for microwave mode was selected by the matrix element for the ``clock'' transition $F_0=2,M_0=0\to F_0=3,M_0=0$ and chosen as $2\bar{U}\simeq 0.05\gamma$. As we can see in Fig. \ref{fig3} the microwave field induces a narrow dip in the imaginary part of $\chi_\perp$ near the precise Raman two-photon resonance $\omega=\omega_c-\Omega$. As was commented at the end of the previous section, the direct coherent Raman conversion of the optical and microwave modes is forbidden because of the selection rules for the electric and magnetic type dipole transitions. The observed transparency window in the spectrum of the susceptibility tensor is another indicator of this effect. Nevertheless such a \textit{microwave field induced transparency} is partly masked in the considered situation since the presence of incoherent scattering mechanisms induces a dissipative behavior in the atomic ground state dynamics, which reduces the coherence and consequently the depth of the observed resonance dip.

In the case of the $\chi_{\parallel}(\omega)$ component, shown in figure \ref{fig4}, the presence of the driving fields is mainly manifestable as a spectral distortion near the $F_0=3\to F=3,2$ transitions and should be associated with the light-induced Zeeman splitting of the upper states. The small suppression of the isotropic part of the susceptibility near $F_0=3\to F=4$ resonant point can be associated with depopulation of level $F_0=3$ initiated by the microwave field.

There are two important kinetic parameters responsible for the light transport. The imaginary part of the sample susceptibility determines the extinction length for the plane wave entering the sample at the $e^{-1}$ level of losses. The inverse extinction length is given by
\begin{equation}
l_{\mathrm{ex}}^{-1}(\omega)=n_0\sigma_{\mathrm{ex}}(\omega)=4\pi k\, \mathrm{Im}\chi(\omega)%
\label{3.2}
\end{equation}
where $n_0$ is a typical local atomic density and for the sake of simplicity we omitted in this estimate its spatial dependence. In an anisotropic sample this quantity can be defined only for a plane wave propagating along specific direction associated with the main reference frame, such that $\chi(\omega)$ in Eq. (\ref{3.2}) can be any of major components of the susceptibility tensor. Otherwise the definition has to depend on the propagation direction and track the polarization properties of light. It is important to point out that in the case of amplification the extinction length accumulates not only losses but also the gain associated with the stimulated Raman scattering. This parameter can be even negative in the case of population inversion and then indicates the distance of $e^{+1}$ amplification.

Another kinetic parameter responsible for the scattering process, and in particular for light trapping, is the scattering length, which is given by
\begin{eqnarray}
l_{\mathrm{sc}}^{-1}(\omega)&=&n_0\sigma_{\mathrm{sc}}(\omega) ,%
\nonumber\\%
\sigma_{\mathrm{sc}}(\omega)&=&\frac{\omega\omega'^3}{c^4}\frac{1}{2F_0+1}\sum_{m'\!,m,\mathbf{e}'}\int\left|\alpha_{\mu'\mu}^{(m'm)}(\omega)e'_{\mu'}e_{\mu}\right|^2d\Omega ,%
\nonumber\\%
\label{3.3}
\end{eqnarray}
where we introduced the scattering cross section $\sigma_{\mathrm{sc}}(\omega)$. Here $\omega,\,\mathbf{e}$ and $\omega',\,\mathbf{e}'$ are the frequencies and polarization vectors of the input and output photons respectively and the integral is over the full scattering angle. Similarly to the extinction length, this quantity critically depends on the polarization direction of the incident photon.

Light diffusion mostly depends on elastic contribution when $\omega=\omega'$ and both $m$ and $m'$ belong the same $F_0=3$ level. With keeping only elastic contribution we can define the characteristic length associated to losses,
\begin{equation}
l_{\mathrm{ls}}^{-1}(\omega)\equiv -l_{\mathrm{g}}^{-1}(\omega)=l_{\mathrm{ex}}^{-1}(\omega)-l_{\mathrm{sc}}^{-1}(\omega) ,%
\label{3.4}
\end{equation}
which evaluates the averaged distance that a photon travels before being lost by an inelastic scattering event. The first equality indicates that in an amplifying medium it is possible to get this quantity negative and then redefine it as the gain length $l_{\mathrm{g}}(\omega)$. In such a gain medium the radiation trapping and amplification mechanisms can overcome the instability point, entering the regime of random Raman laser generation. In the original Letokhov's model of the random lasing phenomenon \cite{Letokhov,Cao}, the instability point obeys the criterion $L^2\gtrsim O(1)l_{\mathrm{sc}}l_{\mathrm{g}}$, where $L$ is the sample size. With atoms, both characteristic lengths can be computed from the atomic response [Eqs. (\ref{3.2}-\ref{3.4})] (see also \cite{FGCK} for more simple approaches with other atomic schemes).

\begin{figure}[tp]
{$\scalebox{0.8}{\includegraphics*{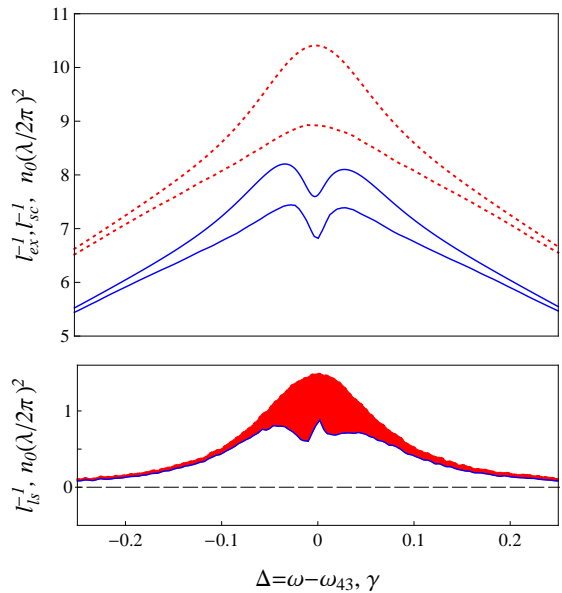}}$ }
\caption{(Color online) Upper panel: extinction length (upper curves) vs. scattering length (lower curves) in the presence of the control optical mode only (red dashed) and in the presence of both the optical and microwave fields (blue solid). The parameters are the same as in Fig. \ref{fig3}. Lower panel: related lengths of losses, see definition (\ref{3.4}). The filled area (below the line corresponding to the absence of pump/microwave excitation) indicates the region where the trapped light can be amplified by the stimulated Raman process due to microwave induced repopulation of atoms on level $F_0=2$.}
\label{fig5}%
\end{figure}%

The typical spectral behavior of the above parameters is shown in Fig. \ref{fig5} for the same external field amplitudes and mutual polarizations as in Figs. \ref{fig3}-\ref{fig4}. The plotted dependencies reveal the  role of the inverse anti-Stokes spontaneous scattering as the main channel of losses. The outcome from this scattering channel rises up significantly due to the Autler-Townes resonance structure created by the external driving fields. Nevertheless the system can be adjusted for observation of the stimulated Raman amplification. The lower panel of Fig. \ref{fig5} shows that atomic repopulation, initiated by the microwave mode, induces  spontaneous Raman emission, which can be additionally stimulated by any light probe propagating through the sample via diffusion. The spectral domain where such Raman amplification partly reduces the losses is indicated by the red-filled area in the plot. In the next section we demonstrate this effect by Monte-Carlo simulations.

By concluding this section let us make one important remark concerning the calculations performed near the microwave induced transparency resonance. The small peak at the center of this resonance appearing in the lower panel of Fig. \ref{fig5} is actually an irrelevant consequence of the performed calculation scheme. In the vicinity of the transparency resonance the instability of the state $F_0=3$, perturbed by the microwave field, manifests itself in that the elastically propagating mode slightly fluctuates near its carrier frequency. This circumstance was commented in section \ref{II.B}. The related spectral distribution, associated with the microwave perturbation, is quite narrow and mostly ignorable, but it is not negligible near the microwave transparency dip. The reason of that is that the interaction process has a long time at the point of microwave induced transparency and the instability of level $F_0=3$ becomes visible effect.

\subsection{Light transport and Raman amplification}\label{Sec.III.B}

In our numerical simulations we consider an atomic cloud with a Gaussian density distribution
\begin{equation}
n(\mathbf{r})=n_0\exp{\left(-\frac{\mathbf{r}^2}{2r_0^2}\right)} ,%
\label{3.5}
\end{equation}
where $n_0$ is the peak density and $r_0$ is the Gaussian radius of the sample. To visualize how light diffusion is modified by the coherent control let us consider a point-like light source located inside the atomic sample and generating an isotropically distributed probe light at frequency $\omega$ near $\omega_{43}$, where losses are reduced by stimulated emission [Fig. \ref{fig5}]. Without driving fields the light would diffusely escape the sample from its boundaries. Then by turning on the optical and microwave modes we can follow the modification of the diffusion process and amplification of the output intensity.

\begin{figure}[tp]
{$\scalebox{0.8}{\includegraphics*{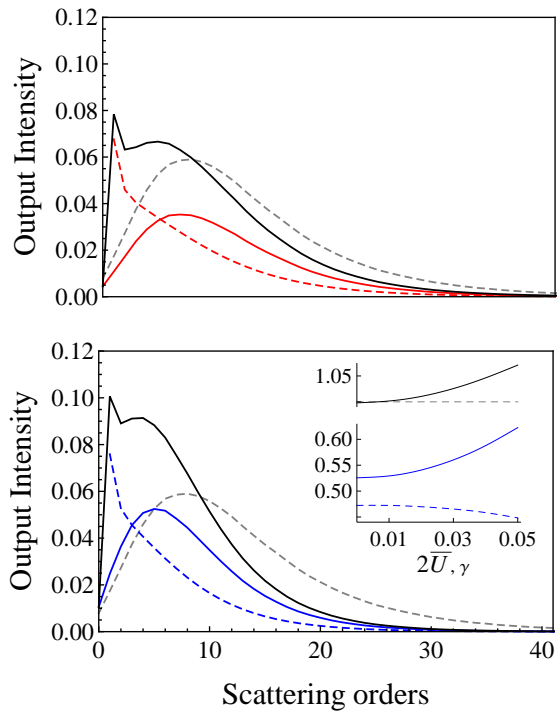}}$ }
\caption{(Color online) Intensity distribution over the scattering orders for a point-like isotropic light source located at the center of the atomic sample. The dashed gray curve in both panels indicates the reference (without driving fields) distribution for light trapping associated with elastic scattering on the closed $F_0=3\to F=4$ transition. The upper panel shows how this distribution is modified by the action of the control optical mode: the red solid and the red dashed curves respectively indicate the distribution for elastic and anti-Stokes inelastic scattering channels; the black solid curve gives the total outcome. The lower panel shows the similar distributions (blue solid for elastic, and blue dashed for inelastic) when both the optical and microwave modes are turned on. The output intensity for the overall outgoing light via both elastic and inelastic channels overcomes the input intensity. The inset shows the total amplification given by the sum over all the scattering orders as a function of the amplitude of the microwave mode.}
\label{fig6}%
\end{figure}

In Fig. \ref{fig6} we show an example of our Monte-Carlo simulations done for an isotropic light source located in the middle of the atomic cloud with optical depth $b_0=\sqrt{2\pi}n_0\sigma_0r_0\sim 10$, where $\sigma_0$ is the resonance cross section for the closed $F_0=3\to F=4$ transition. The source frequency is selected in the amplification region shown in Fig. \ref{fig5} and the graphs reproduce typical behavior similar for all the frequencies of this region. The graphs demonstrate how the output intensity generated in the full solid angle by the light source is distributed over the different orders of light scattering i. e. over partial contributions of the Bethe-Salpeter expansion, such that the total outcome is given by the sum over all the partial contributions.

The upper panel in figure \ref{fig6} demonstrates how the optical control mode affects the process. The gray dashed curve in both panels indicates the reference distribution of the outgoing light associated with elastic scattering on the closed $F_0=3\to F=4$ transition such that the sum over all orders is conventionally normalized to the unit input intensity. The contribution associated with zero order means the light escaping the sample via elastic channel without any scattering. Without any driving fields the anti-Stokes Raman scattering via the $F_0=3\to F=3,2 \to F_0=2$ transitions makes a negligible contribution into any random path associated with the light trapping process. But turning on the optical mode enhanced this anti-Stokes channel, which includes two contributions: a dynamical $\Lambda$-type interaction between the control and probe modes and a spontaneous part. The dynamical interaction is a reversible process, which eventually modifies the diffusion transport of the elastically scattered light, as shown by the red solid curve in the plot, but does not create any loss. On the contrary, the spontaneous inelastic anti-Stokes scattering, tracked by the red dashed curve, can be interpreted as a mechanism of losses redistributing the fractions of light from its original elastic trapping path. It is important to recognize that the inelastic scattering happens only one time as the last event removing the photon from the sample, whereas multiple scattering happens on the $F_0=3\to F=4$ line many times. So many elastic events preceding the last inelastic one are accumulated in the definition of the scattering order for the anti-Stokes graphs.  The sum of both elastic and inelastic scattering channels, indicated by black solid curve, obeys the conservation law such that the total intensity is unchanged and equal to the input intensity.

Turning on the microwave field activates the Raman amplification process shown in Fig.~\ref{fig1}. The lower panel of Fig.~\ref{fig6} clearly demonstrates the stimulated amplification of the outgoing light. The light emerging the sample via the elastic channel, indicated by the blue solid curve, is stronger than in the presence of the optical mode only (red solid line in the upper panel). The losses associated with inverse anti-Stokes scattering channel, shown by the blue dash line, are also slightly reduced. The output intensity for the overall outgoing light via both elastic and inelastic channels is about five percents larger than the input intensity. In addition in inset we reproduce the dependence of the total amplification given by the sum over all the scattering orders as a function of the amplitude of the microwave mode.

\begin{figure}[tp]
{$\scalebox{0.8}{\includegraphics*{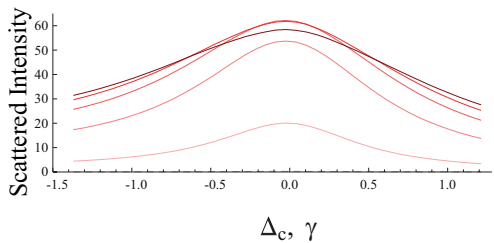}}$ }
\caption{(Color online) Intensity of the light scattered at the angle $\theta=0^0$ [Fig.~\ref{fig2}]. The control mode frequency is scanned near the $F_0=2\to F=4$ forbidden transition such that $\Delta_c=\omega_c-\omega_{42}-\Delta_L$, where $\Delta_L$ is the light shift of level $F_0=2$ induced by this mode. The scattered intensity (number of photons) is given by Eq.~(\ref{2.2}) calculated for the same driving field parameters as in Fig.~\ref{fig5}. The curve thickness in the plotted graphs is associated with optical thickness changed from lower to higher values $b_0=1,5,10,15,20$. For this observation angle only elastic scattering contributes to the emitted intensity.}
\label{fig7}%
\end{figure}

\begin{figure}[tp]
{$\scalebox{0.8}{\includegraphics*{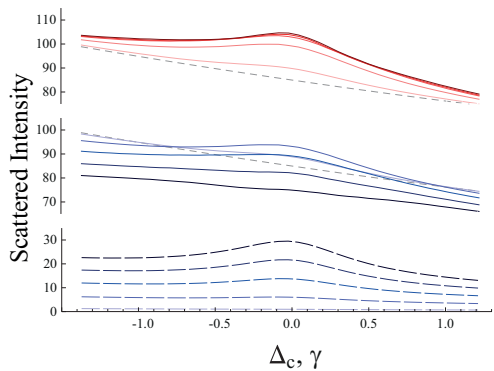}}$ }
\caption{(Color online) Same as Fig.~\ref{fig7} but for the observation angle $\theta=45^0$. The lower panel shows the contribution of the inverse anti-Stokes scattering, the middle panel is the contribution of elastic scattering, and the upper panel gives the sum for both the channels. The curve thickness in the plotted graphs is associated with optical thickness changed from lower to higher values $b_0=1,5,10,15,20$.}
\label{fig8}%
\end{figure}

\begin{figure}[tp]
{$\scalebox{0.8}{\includegraphics*{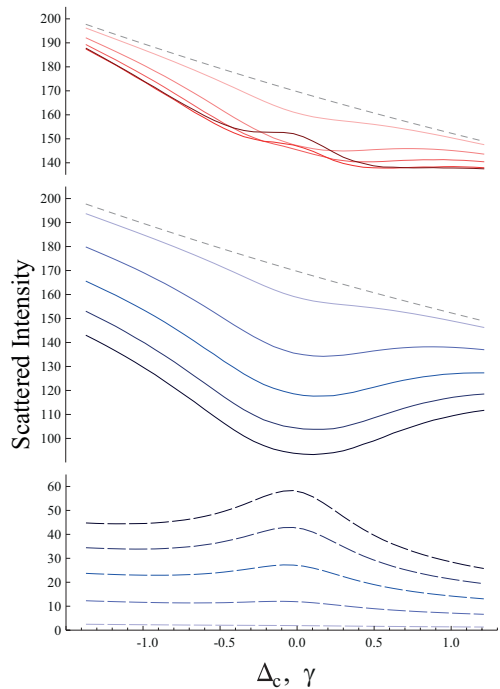}}$ }
\caption{(Color online) Same as Fig.~\ref{fig8} but for the observation angle $\theta=90^0$.}
\label{fig9}%
\end{figure}

Let us now consider the light dynamics initiated by the basic excitation process shown in Fig.~\ref{fig1} via calculation of the correlation function (\ref{2.1}) and its spectral component (\ref{2.2}) for the scattered light. In Figs.~\ref{fig7}-\ref{fig9} we show the spectral variation for the intensity of the light originally emitted by the Raman process and further scattered by the atomic ensemble under the trapping conditions. The frequency of the incident light (control mode) is scanned near the forbidden $F_0=2\to F=4$ transition and the optical thickness of the atomic sample is varied from low to high values $b_0=1,5,10,15,20$. The seeding contribution is generated by the second diagram in the right side of equation (\ref{2.5}) and is associated with spontaneous Raman emission. This process is distributed inside the sample and generates non-monochromatic light with narrow but finite spectral profile as explained in the previous section. This source term has its graph image similar to the amplification term in the self-energy part of the retarded-type photon propagator calculated in appendix \ref{Appndx_A}. As a consequence its analytical structure is more or less identical to the gain contribution in the extinction length (\ref{3.2}) and the spontaneous Raman emission is distributed in the same spectral domain as shown by filled area in figure \ref{fig5}. Further details concerning the source term and the iterative solution of the Bethe-Salpeter equation are discussed in appendix \ref{Appndx_B}. In our Monte-Carlo simulations we follow the scattering response at the particular frequency inside the amplification area of figure \ref{fig5} and for the same driving field parameters. The three plots of figures \ref{fig7}-\ref{fig9} respectively relate to three different observation angles $\theta=0^0,\,45^0,\,90^0$ defined from the control mode polarization direction [Fig.~\ref{fig2}].  The microwave mode follows the light shift of the level $F_0=2$ and is always tuned between the lower hyperfine sublevels so that it provides equal conditions for atomic repopulation for all the control mode detunings.

As follows from the plotted graphs even in the spectral domain far off-resonant from the $F_0=2\to F=4$ resonance the emitted radiation undergoes a sequence of scattering on the Autler-Townes resonance structure. In the optically thick system part of the light escapes the sample via the inverse anti-Stokes scattering channel and part via the elastic channel. Because of the polarization anisotropy of the entire process in the dense system the scattered radiation is spatially redistributed and emerges the sample preferably in the direction orthogonal to the control mode polarization. When the control mode scans the vicinity of the $F_0=2\to F=4$ transition it stimulates Raman amplification of the trapped light, which can be seen in both the elastic and inelastic channels. For the specific observation angle $\theta=0^0$, for which there is no direct scattering on the Autler-Townes resonance, we can clearly observe Raman-stimulated amplification of light diffusion via the elastic channel, see Fig.~\ref{fig7}. But in the general case the trapped light can be converted into light emission at the frequency of the incident mode and this emission has a signature of the stimulated Raman amplification as well. Let us point out here that in our simulations we ignored contribution of the direct Rayleigh scattering of the control mode which would give a flat background to all the plotted dependencies.

Summarizing the results of our Monte-Carlo simulations we have demonstrated that the Raman-stimulated amplification of the light trapped on the closed $F_0=3\to F=4$ transition is strongly affected by the losses via the inverse anti-Stokes scattering channel. This process counteracts the lasing mechanism but enhances the overall emission associated with the redistribution of the control light via all the scattering channels.

\subsection{Towards the instability conditions}\label{Sec3C}

The above calculations were presented for the situation when only a small number of atoms are repopulated onto the pump level $F_0=2$. That allowed us to make a detailed Monte Carlo simulations of the randomized Raman process with keeping all the contributing orders of the multiple scattering. Such a configuration is evidently not optimal in order to approach the instability point and visualize the random lasing effect, which in our case would mean divergence of the iterative scheme applied to the Bethe-Salpeter equation. The sufficient condition for the instability poses that the length of losses $l_{\mathrm{ls}}$, given by Eq.~(\ref{3.4}), should be negative and transformed to a gain length. The simplest way to reach this regime is to provide a population inversion in the dielectric susceptibility given by Eq.~(\ref{A.8}). But that would be difficult to do with coherent microwave coupling and there are other  possibilities, which could help to reduce the problem with light losses.

If the atomic sample is dense enough to contain more than one atom in the volume scaled by the radiative wavelength then the near field interaction and cooperative dipole dynamics could enhance the effective coupling with the emitted modes. As known from recent studies of highly disordered three dimensional atomic systems such effects crucially modify the light transport dynamics, see \cite{Akkermans,SKKH}. In certain conditions the effective coupling with the populated sublevels could demonstrate a signature of cavity behavior for the trapped light. In this case the spontaneous Raman scattering would emit the photons in the sample and the effective coupling strength for this process would be enhanced because of the local field effects. Further amplification of the spontaneously emitted light would be supported by the presence of a local random cavity built with the atomic local environment and the entire process could attain the lasing threshold. Note that such cooperative scattering effect are included in the random-laser theory of Goetschy and Skipetrov \cite{Goetschy:2011}. Let us also point out that diverging regimes in a simulation of elastic multiple scattering on Rayleigh particles were discussed many years ago in \cite{SinghamBoren}.

There is another possibility to eliminate the losses via preparing spatially-inhomogeneous energy structure and population distribution of the hyperfine sublevels in the atomic ensemble. A relevant experimental design has been recently proposed in M. Havey \cite{Havey} and implies the idea of controllable light shift of only one hyperfine energy level for the atoms located in a spatially selected cylindrical volume inside the cloud. If in our case we assumed the level $F_0=2$ as light shifted and organized the population inversion for all the atoms in the selected volume onto this level (for example, with $\pi$-type microwave pulse) then these atoms could form an active medium for the photon emission. Indeed, as follows from the transition scheme [Fig.~\ref{fig1}], the control mode would create the photon emission on $F=4\to F_0=3$ only for the atoms inside the selected volume and the interaction would be off-resonant with both the microwave and control fields for the other atoms of the ensemble. Then the atoms outside of the active volume would trap the emitted light and play the role of a soft cavity redirecting the light in a quasi one-dimensional propagation channel associated with that volume. That could lead to instability if each spontaneously emitted photon would have a diffusion path long enough for stimulation of extra photon emission while it propagates through the channel.


\section{Summary}

In the paper we considered the Raman process developing in the D$_2$ hyperfine manifold of alkali-metal atom and initiated by simultaneous action of linearly polarized coherent optical and microwave modes in conditions such that the emitted light is trapped by multiple scattering in the atomic medium. Initial coherent excitation can initiate only incoherent Raman emission because the direct coherent three-wave conversion of the driving optical and microwave modes is forbidden by the considered symmetry of the Lambda-type transition . The light frequency of the spontaneously emitted mode is tuned resonantly to the closed $F_{0\max}\to F_{\max}$ transition (which is $F_0=3\to F=4$ in our calculations for ${}^{85}$Rb) such that the emitted light is expected to be trapped by the medium via a diffusion process due to elastic scattering. However we obtained that its diffusion propagation is also strongly affected by inelastic inverse anti-Stokes scattering on the Autler-Townes resonance created by the control optical mode, which induces leakage from any elastic propagation channel.

Our Monte-Carlo simulations of the entire process in various conditions demonstrates some specificity in the light transport behavior. There is a narrow spectral window where the presence of the microwave mode eliminates the Lambda-coupling of the diffused light with the control mode and yields only elastic light diffusion via its successive scattering on the closed transition. But in general situation the trapped light can additionally stimulate the Raman emission inside the sample via the Lambda-coupling and the initiated process of Raman amplification is stronger as more atoms are repopulated in the $F_{0\min}$ hyperfine sublevel of the ground state ($F_0=2$ in the case of ${}^{85}$Rb). As we did not enter the regime of large gain (negative $l_\mathrm{ls}$) the losses associated with the scattering on the Autler-Townes resonance counteract the approach to the random lasing threshold.

Nevertheless we have pointed out possible modifications of the considered scheme towards the conditions where the amplification process could attain the instability point and be transformed to the random lasing regime. As we expect, this could be achieved in a dense and highly disordered atomic system with spatially inhomogeneous energy structure and distribution of atoms over the hyperfine sublevels. In some specific configurations the local field effects could enhance the atomic interaction with the trapped modes or it could be possible to spatially separate the process of light amplification from its diffusion dynamics.

\section*{Acknowledgments}

We thank Igor Sokolov and Mark Havey for many fruitful discussions. This work was supported by the CNRS-RFBR collaboration (CNRS 6054 and RFBR 12-02-91056), by RFBR-13-02-00944 and the ANR (ANR-06-BLAN-0096). D.V.K. would like to acknowledge support from the External Fellowship Program of the Russian Quantum Center (Ref. Number 86). L.V.G. would like to acknowledge support from the Foundation ``Dynasty''.

\appendix

\section{Photon propagator}\label{Appndx_A}

\subsection{Definition}

If a weak coherent mode propagates through the atomic sample its amplitude is modified in accordance with the following graph equation
\begin{eqnarray}
\lefteqn{\langle \hat{A}_{\mu}^{(+)}(\mathbf{r},t)\rangle=\left\langle S^{\dagger}%
T\!\left(\!S\hat{A}_{\mu}^{(0,+)}(\mathbf{r},t)\!\right)\right\rangle\Rightarrow}%
\nonumber\\%
&&\scalebox{1.0}{\includegraphics*{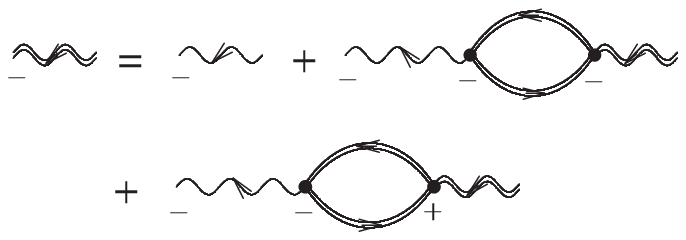}}%
\label{A.1}%
\end{eqnarray}
The self-energy part of this diagram performs the retarded-type polarization operator, and the internal double lines express the ``dressed'' atomic propagators, which have accumulated all the interaction processes. Each vertex has either positive or negative sign, depending on the interaction terms from either expansion of $S^\dagger$ (pointed by ``+'') or $S$ (pointed by ``-''). The graph equation encodes the macroscopic wave equation for the probe propagating in the bulk medium. The fundamental solution of equation (\ref{A.1}) performs the retarded-type Green's function (photon propagator), which was formally defined by Eq.~(\ref{2.6}).

For the stationary conditions the retarded-type propagator (\ref{2.6}) depends only on the difference of its time arguments $\tau=t_1-t_2>0$. Then it can be equivalently introduced by its Fourier transform
\begin{equation}
{\cal D}_{\mu_1\mu_2}^{(R)}(\mathbf{r}_1,\mathbf{r}_2;\omega)=%
\int_0^\infty\!\! d\tau\ \mathrm{e}^{i\omega\tau}\ %
{\cal D}_{\mu_1\mu_2;+}^{(R)}(\mathbf{r}_1,\mathbf{r}_2;\tau),%
\label{A.2}%
\end{equation}
which is considered for $\omega>0$, i.e., for its positive frequency part. This function performs the fundamental solution of the following macroscopic wave equation
\begin{eqnarray}
\lefteqn{\left[\triangle_1+\frac{\omega^2}{c^2}\right]%
{\cal D}_{\mu_1\mu_2}^{(R)}(\mathbf{r}_1,\mathbf{r}_2;\omega)}%
\nonumber\\%
&&+\;\sum_{\mu}\int\!\!d^3r\ {\cal P}_{\mu_1\mu}^{(R)}(\mathbf{r}_1,\mathbf{r};\omega)\,%
{\cal D}_{\mu\mu_2}^{(R)}(\mathbf{r},\mathbf{r}_2;\omega)%
\nonumber\\%
&&=4\pi\!\hbar\,\delta_{\mu_1\mu_2}^{\perp}(\mathbf{r}_1-\mathbf{r}_2),%
\label{A.3}%
\end{eqnarray}
where the transverse projection of the polarization operator, entering the Dyson equation, can be expressed by the dielectric susceptibility of the medium via the following integral relation
\begin{eqnarray}
{\cal P}_{\mu_1\mu}^{(R)}(\mathbf{r}_1,\mathbf{r};\omega)&=&%
4\pi\frac{\omega^2}{c^2}\sum_{\mu'\mu''}\int\!\!d^3r'\ \delta_{\mu_1\mu'}^{\perp}(\mathbf{r}_1-\mathbf{r}')%
\nonumber\\%
&&\times\,\chi_{\mu'\mu''}\!(\mathbf{r}',\omega)\,\delta_{\mu''\mu}^{\perp}(\mathbf{r}'-\mathbf{r})%
\label{A.4}%
\end{eqnarray}
where
\begin{eqnarray}
\lefteqn{\hspace{-0.5cm}\delta_{\mu\mu'}^{\perp}(\mathbf{r}-\mathbf{r}')=\int\!\! \frac{d^3k}{(2\pi)^3}%
\left[\delta_{\mu\mu'}-\frac{k_\mu k_{\mu'}}{k^2}\right]\,\mathrm{e}^{i\mathbf{k}\cdot(\mathbf{r}-\mathbf{r}')}}%
\nonumber\\%
&=&\delta_{\mu\mu'}\delta(\mathbf{r}-\mathbf{r}')+%
\frac{1}{4\pi}\frac{\partial^2}{\partial x_{\mu}\partial x_{\mu'}}\frac{1}{|\mathbf{r}-\mathbf{r}'|}%
\label{A.5}%
\end{eqnarray}
and the same transverse delta function contributes in the r.h.s. of the basic equation (\ref{A.3}).

\subsection{The dielectric susceptibility}

The dielectric susceptibility of the medium can be introduced by the Kubo formula \cite{LaLfIX}
\begin{equation}
\chi_{\mu\mu'}\!(\mathbf{r},t;\mathbf{r}',t')%
=\frac{i}{\hbar}\left\langle \left[P_{\mu}(\mathbf{r},t),P_{\mu'}(\mathbf{r}',t')\right]\right\rangle\theta(t-t'),%
\label{A.6}%
\end{equation}
where $P_{\mu}(\mathbf{r},t)$ is the local polarization of the sample given by the microscopic distribution of the atomic dipole moments. For motionless system of atomic dipoles in stationary conditions and for spatially inhomogeneous configuration the susceptibility can be factorized in the following product
\begin{equation}
\chi_{\mu\mu'}\!(\mathbf{r},t;\mathbf{r}',t')\ =\ \chi_{\mu\mu'}\!(\mathbf{r};t-t')\;\delta(\mathbf{r}-\mathbf{r}').%
\label{A.7}%
\end{equation}
Here the presence of the $\delta$-function indicates that spatial dispersion vanishes for motionless point-like dipole system. The retarded-type polarization operator, given by its diagram definition [Eq.~(\ref{A.1})], shows that the susceptibility $\chi_{\mu\mu'}\!(\mathbf{r};t-t')$ can be expanded as a product of the atomic Green's functions of the ground and excited states. In turn the dynamics of the atomic Green's function consists of two mechanisms: 1) either slow variation of the density matrix in time with approaching the steady state level; 2) their degradation as a function of $\tau=t-t'$ in accordance with the quantum regression theorem \cite{CohenTannoudji}. The latter is performed by the retarded and advanced atomic Green's functions of the upper and ground states, which contribute in the graph representation of the polarization operator.

After Fourier transformation the susceptibility tensor is given by
\begin{eqnarray}
\lefteqn{\chi_{\mu\mu'}(\mathbf{r},\omega)=%
\frac{i}{\hbar}\sum_{m,m'}\sum_{n,n'}n_0(\mathbf{r})\left(d_{\mu}\right)_{mn}%
\left(d_{\mu'}\right)_{n'm'}}%
\nonumber\\%
&&\times\left[\sum_{m''}\int_{-\infty}^{\infty}\frac{dE}{2\pi\hbar}{\cal G}_{nn'}^{(R)}(E)\bar{\rho}_{m'm''}{\cal G}_{m''m}^{(A)}(E-\hbar\omega)\right.%
\nonumber\\%
&&\left.-\sum_{n''}\int_{-\infty}^{\infty}\frac{dE}{2\pi\hbar}{\cal G}_{nn''}^{(R)}(E)\bar{\rho}_{n''n'}{\cal G}_{m'm}^{(A)}(E-\hbar\omega)\right]\!\!,%
\label{A.8}%
\end{eqnarray}
where we introduced the local density of atoms $n_0(\mathbf{r})$ at point $\mathbf{r}$ and steady state components of the density matrix for the ground state $\bar{\rho}_{m'm''}$ (both indices belong to the same $F_0=3$ hyperfine sublevel) and for the excited states $\bar{\rho}_{n''n'}$ (where $n''$ and $n'$ can be any allowed substates of the upper level hyperfine manifold).

The derived expression (\ref{A.8}) is quite general and incorporates all the nonlinear ``dressing'' corrections associated with the control and microwave modes in the retarded-type ${\cal G}_{nn'}^{(R)}(E)$ and advanced-type ${\cal G}_{m'm}^{(A)}(E)$ atomic propagators. But it ignores any nonlinear contributions associated with the weak probe mode. The most valuable is that the susceptibility tensor is expressed in a form where density matrix elements can be independently calculated via Bloch-type master equations (see appendix \ref{Appndx_C}), but they also can be considered as external parameters affecting the process of light transport.

The crucial elements in Eq.~(\ref{A.8}) are the atomic propagators. For the excitation scheme described in section \ref{II.A}, where both control and microwave modes have the same linear polarization, these propagators can be found in a closed analytical form. Their diagram analysis and evaluation includes subsequent "dressing" by spontaneous decay and by coherent coupling with control and microwave modes. The retarded-type atomic propagator has the graph definition given by the difference of the $T$-ordered and $N$-ordered products of the atomic second quantized operators and obeys the following Dyson equation
\begin{equation}
\scalebox{0.9}{\includegraphics*{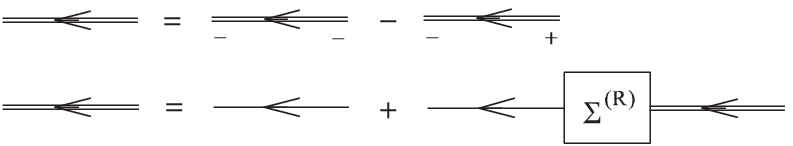}}%
\label{A.9}%
\end{equation}
where its self-energy part is given by
\begin{equation}
\scalebox{0.8}{\includegraphics*{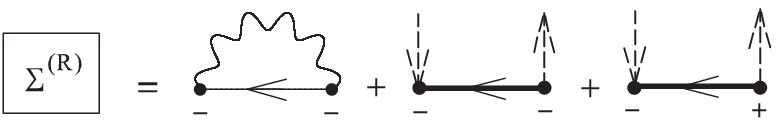}}%
\label{A.10}%
\end{equation}
Here the first contribution is responsible for the standard spontaneous decay, other terms express the interactions with the control field and the dashed arrow-ended lines show coupling of the upper states with the ground state sublevel $F_0=2$. In these diagrams the internal solid lines perform the Green's functions of the atomic states belonging level $F_0=2$ and dressed by the interaction with the microwave mode only. Both the atomic line contributed into the second and third term of (\ref{A.10}) can be incorporated into a retarded-type atomic propagator, which obeys the following Dyson equation
\begin{equation}
\scalebox{0.9}{\includegraphics*{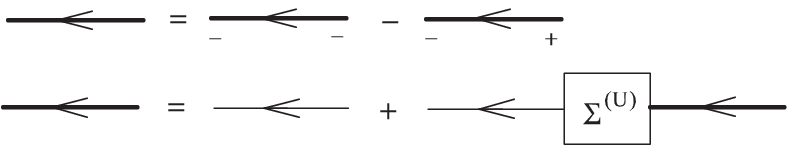}}%
\label{A.11}%
\end{equation}
which self-energy part includes only interaction with the microwave mode
\begin{equation}
\scalebox{0.6}{\includegraphics*{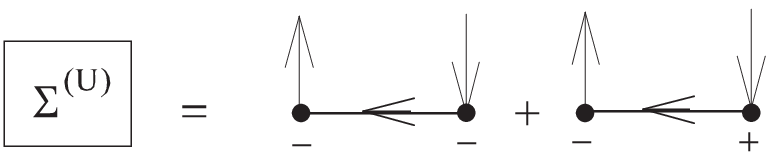}}%
\label{A.12}%
\end{equation}
where the thin arrow-ended lines indicate the coupling of the ground state hyperfine sublevels via the microwave mode. The self-energy part of the latter equation contains only undisturbed vacuum atomic lines, such that its solution can be found in analytical form. The obtained result should be substituted into the self-energy block of (\ref{A.10}) and then transform the Dyson equation (\ref{A.9}) to the closed and solvable form.

Further derivation is quite straightforward and omitting its details we show here the final result only. Propagator ${\cal G}_{nn'}^{(R)}(E)$ can be found as solution of the following system of algebraic equations
\begin{eqnarray}
\lefteqn{\left[E-E_n+i\hbar\frac{\gamma}{2}\right]{\cal G}_{nn'}^{(R)}(E)\ -\ \sum_{n''}\ V_{n\tilde{m}}V_{n''\tilde{m}}^{*}}%
\nonumber\\%
&&\times\left\{E-\hbar\omega_c-E_{\tilde{m}}-\frac{|U_{m\tilde{m}}|^2}{E-\hbar(\omega_c-\Omega)-E_{m}+i0}\right\}^{-1}%
\nonumber\\%
&&\phantom{\frac{|u|^2}{E-\hbar\omega}}\times\ {\cal G}_{n''n'}^{(R)}(E)=\ \hbar\,\delta_{nn'},%
\label{A.13}
\end{eqnarray}
where $V_{n\tilde{m}}$ and $V_{n''\tilde{m}}$ are the matrix elements of interaction with the control mode, $E_n$ is the energy of the upper state, $\gamma$ is its spontaneous decay rate, $\omega_c$ is the frequency of the control mode and $\Omega$ is the microwave frequency. For linearly polarized driving light, in accordance with the selection rules for the transition matrix elements, any excited state $|n\rangle,|n''\rangle$ contributing in these equations can be coupled only to one specific Zeeman state $|\tilde{m}\rangle$ belonging to the ground state hyperfine sublevel $F_0=2$. In turn, for the interaction with the linearly polarized microwave field, described by its matrix elements $U_{m\tilde{m}}$, the state $|\tilde{m}\rangle$ can be coupled only with one specific Zeeman state $|m\rangle$ belonging to the hyperfine sublevel $F_0=3$.

The ground state advanced-type propagator ${\cal G}_{m'm}^{(A)}(E)$ can be similarly calculated and is given by
\begin{eqnarray}
\lefteqn{\hspace{-0.5cm}{\cal G}_{m'm}^{(A)}(E)=\delta_{m'm}\;%
\hbar\left\{E-E_{m}-|U_{m\tilde{m}}|^2%
\left[E-\hbar\Omega-E_{\tilde{m}}\phantom{\sum_n\frac{|V|^2}{E}}\right.\right.}%
\nonumber\\%
&&\left.\left.-\sum_n\frac{|V_{n\tilde{m}}|^2}{E+\hbar(\omega_c-\Omega)-E_{n}-i\hbar\gamma/2}\right]^{-1}\right\}^{-1}%
\label{A.14}
\end{eqnarray}
and has diagonal structure.

\subsection{The radiation zone asymptote}

The solution of equation (\ref{A.3}) physically describes the radiation response generated by a point-like atomic dipole, which is considered as a wave source. Attenuation of the scattered spherical wave is scaled by the extinction length $l_0$, which in a dilute system is always much longer than the wavelength $\lambdabar$. Thus for practical applications it is enough to find a long distant (radiation zone) asymptote for ${\cal D}_{\mu_1\mu_2}^{(R)}(\mathbf{r}_1,\mathbf{r}_2;\omega)$ when $|\mathbf{r}_1-\mathbf{r}_2|\sim l_0\gg\lambdabar$. This can be done with the phase integral technique.

The relevant distribution volume in the integral over $\mathbf{r}$ in Eq.~(\ref{A.3}) is in a small vicinity around the point $\mathbf{r}_1$, which can be bounded by a box of a few $\lambda$. The atomic density distribution as well as the sample susceptibility can be considered as homogeneous functions inside this volume. That allows the Green's function to be factorized in the product
\begin{equation}
{\cal D}_{\mu_1\mu_2}^{(R)}(\mathbf{r}_1,\mathbf{r}_2;\omega)\approx%
-\hbar X_{\mu_1\mu_2}(\mathbf{r}_1,\mathbf{r}_2;\omega)%
\frac{\exp\left[ik|\mathbf{r}_1-\mathbf{r}_2|\right]}{|\mathbf{r}_1-\mathbf{r}_2|},%
\label{A.15}%
\end{equation}
where the slow varying amplitude $X_{\mu_1\mu_2}(\mathbf{r}_1,\mathbf{r}_2;\omega)$ obeys the following differential equation
\begin{eqnarray}
\frac{\partial}{\partial z_1}X_{\mu_1\mu_2}\!(\mathbf{r}_1,\mathbf{r}_2;\omega)\!\!&=&\!\!%
\frac{2\pi i\omega}{c}\!\sum_{\mu}\chi_{\mu_1\mu}\!(\mathbf{r}_1,\omega)%
X_{\mu\mu_2}\!(\mathbf{r}_1,\mathbf{r}_2;\omega)%
\nonumber\\%
X_{\mu_1\mu_2}(\mathbf{r}_1,\mathbf{r}_2;\omega)&\to&\delta_{\mu_1\mu_2}^{\perp}%
\ \ \ \mathrm{at}\ \ \mathbf{r}_1\to\mathbf{r}_2%
\label{A.16}%
\end{eqnarray}
This equation is written in a special reference frame where the $z$-axis is directed along the ray from point $\mathbf{r}_2$ to point $\mathbf{r}_1$. In the radiation zone the ray can be associated with the locally plane wave with the wave vector $\mathbf{k}=k(\mathbf{r}_1-\mathbf{r}_2)/|\mathbf{r}_1-\mathbf{r}_2|$. In turn the polarization components $\mu_1,\mu_2,\mu$ belong to the plane $x,y$, which is transverse to the propagation direction. The Kronecker symbol $\delta_{\mu_1\mu_2}^{\perp}$ is now transverse in the real (not reciprocal) space. Thus the equation (\ref{A.16}) describes the transformation of the complex amplitude of a light plane wave propagating in the space from point $\mathbf{r}_2$ to point $\mathbf{r}_1$.

The dielectric susceptibility tensor contributes to Eq.~(\ref{A.16}) by its projection onto the wavefront $x,y$-plane. Because the helicity convention is more preferable for such frame it is more natural to define the susceptibility components in the angular momentum basis set. This can be done by the following basis expansion $\mathbf{e}_{0}=\mathbf{e}_{z}$, $\mathbf{e}_{\pm 1}=\mp (\mathbf{e}_{x}\pm i\mathbf{e}_{y})/\sqrt{2}$ with respect to the Cartesian components and it requires co/contravariant notation in definition of the tensors indices \cite{VMK}. Let us denote the transverse components of the susceptibility tensor in the angular momentum representation as $\tilde{\chi}_{q_1}{}^{q_2}(\ldots)$, where $q_1,q_2=\pm 1$.
In the considered case, when the susceptibility tensor in its main frame is described by two major components [Eq.~(\ref{3.1})], the required transformation is given by
\begin{eqnarray}
\tilde{\chi}_{+1}{}^{+1}(\mathbf{r},\omega)\!\!&=&\!\!\tilde{\chi}_{-1}{}^{-1}(\mathbf{r},\omega)%
\nonumber\\%
&=&\!\!\frac{1+\cos^2\beta}{2}\,\chi_{\bot}(\mathbf{r},\omega)+\frac{\sin^2\beta}{2}\,\chi_{\parallel}(\mathbf{r},\omega),%
\nonumber\\%
\tilde{\chi}_{\mp 1}{}^{\pm 1}(\mathbf{r},\omega)\!\!&=&\!\!\mathrm{e}^{\pm 2i\gamma}\,\frac{\sin^2\beta}{2}\left[\chi_{\bot}(\mathbf{r},\omega)-\chi_{\parallel}(\mathbf{r},\omega)\right]\!,%
\label{A.17}
\end{eqnarray}
where $\alpha$ (not contributing), $\beta, \gamma$ are the standard Euler angles describing the rotational transformation from the $X,Y,Z$ main frame to the $x,y,z$ frame associated with a light ray.

Then $\tilde{\chi}_{q_1}{}^{q_2}(\ldots)$ is a $2\times 2$ matrix which can be expanded in the set of Pauli matrices $\overrightarrow{\sigma}=(\sigma_{\mathrm{x}},\sigma_{\mathrm{y}},\sigma_{\mathrm{z}})$ as follows
\begin{equation}
\tilde{\chi}_{q_1}{}^{q_2}(\mathbf{r},\omega )=\chi_0(\mathbf{r},\omega)%
\delta _{q_1}{}^{q_2}+%
\left( \overrightarrow{\chi}(\mathbf{r},\omega)\cdot\overrightarrow{\sigma }\right)_{q_1}{}^{q_2}%
\label{A.18}%
\end{equation}%
We conventionally subscribe the Pauli matrices by ``$\mathrm{x},\,\mathrm{y},\,\mathrm{z}$'' indices (in Roman style) and this symbolic vector notation should not be confused with the Cartesian frame notation. In this expansion the upper row and left column of Pauli matrices are associated with the $+1$ index and the respective lower row and right column with the $-1$ index. The expansion coefficients $\chi_{0}(\mathbf{r},\omega)$ and $\overrightarrow{\chi }(\mathbf{r},\omega )$ can be found for each particular tensor of the dielectric susceptibility via the above rotational transformation from laboratory to the local reference frame. In the considered case they are given by
\begin{eqnarray}
\chi_0(\mathbf{r},\omega)&=&\frac{1+\cos^2\beta}{2}\,\chi_{\bot}(\mathbf{r},\omega)+\frac{\sin^2\beta}{2}\,\chi_{\parallel}(\mathbf{r},\omega),%
\nonumber\\%
\chi_\mathrm{x}(\mathbf{r},\omega)&=&\cos 2\gamma\,\frac{\sin^2\beta}{2}\left[\chi_{\bot}(\mathbf{r},\omega)-\chi_{\parallel}(\mathbf{r},\omega)\right],%
\nonumber\\%
\chi_\mathrm{y}(\mathbf{r},\omega)&=&\sin 2\gamma\,\frac{\sin^2\beta}{2}\left[\chi_{\bot}(\mathbf{r},\omega)-\chi_{\parallel}(\mathbf{r},\omega)\right],%
\nonumber\\%
\chi_\mathrm{z}(\mathbf{r},\omega)&=&0,%
\label{A.19}
\end{eqnarray}
where the major components are specified by Eq.~(\ref{3.1}).

Then the solution of equation (\ref{A.16}) can be expressed by the above parameters. Let us enumerate the polarization components $\mu_1,\mu_2$ by $1$ ($x$-axis) and $2$ ($y$-axis), then the solution is given by
\begin{eqnarray}
X_{11}(\mathbf{r}_{2},\mathbf{r}_{1};\omega )\!\!&=&\!\!%
e^{i\phi _{0}(\mathbf{r}_{2},\mathbf{r}_{1})}%
\left[ \cos \phi (\mathbf{r}_{2},\mathbf{r}_{1})-i\sin \phi (\mathbf{r}_{2},\mathbf{r}_{1})n_{\mathrm{x}}\right],%
\nonumber \\%
X_{22}(\mathbf{r}_{2},\mathbf{r}_{1};\omega )\!\!&=&\!\!%
e^{i\phi _{0}(\mathbf{r}_{2},\mathbf{r}_{1})}%
\left[ \cos \phi (\mathbf{r}_{2},\mathbf{r}_{1})+i\sin \phi (\mathbf{r}_{2},\mathbf{r}_{1})n_{\mathrm{x}}\right],%
\nonumber \\%
X_{12}(\mathbf{r}_{2},\mathbf{r}_{1};\omega )\!\!&=&\!\!%
e^{i\phi_{0}(\mathbf{r}_{2},\mathbf{r}_{1})}\,i\sin \phi (\mathbf{r}_{2},\mathbf{r}_{1})%
\left(n_{\mathrm{y}}+i\,n_{\mathrm{z}}\right),%
\nonumber\\%
X_{21}(\mathbf{r}_{2},\mathbf{r}_{1};\omega )\!\!&=&\!\!%
e^{i\phi_{0}(\mathbf{r}_{2},\mathbf{r}_{1})}\,i\sin \phi (\mathbf{r}_{2},\mathbf{r}_{1})%
\left(n_{\mathrm{y}}-i\,n_{\mathrm{z}}\right),%
\label{A.20}%
\end{eqnarray}%
where
\begin{eqnarray}
\phi _{0}(\mathbf{r}_{2},\mathbf{r}_{1})&=&\frac{2\pi \omega }{c}%
\int_{\mathbf{r}_{1}}^{\mathbf{r}_{2}}\chi_0(\mathbf{r},\omega )ds,%
\nonumber\\%
\phi (\mathbf{r}_{2},\mathbf{r}_{1})&=&\frac{2\pi \omega }{c}%
\int_{\mathbf{r}_{1}}^{\mathbf{r}_{2}}\chi (\mathbf{r},\omega )ds,%
\label{A.21}%
\end{eqnarray}%
performs the phase integrals along the ray $s$ linking the points $\mathbf{r}_1$ and $\mathbf{r}_2$ such that $\mathbf{r}=\mathbf{r}(s)$ in the integrand.

In these integrals $\chi(\mathbf{r},\omega)$ is the complex ``length'' of the symbolic vector $\overrightarrow{\chi}(\mathbf{r},\omega)$ and $\overrightarrow{n}(\omega )$ is its complex  ``director'', which are given by
\begin{eqnarray}
\chi^2(\mathbf{r},\omega)&=&%
\chi_\mathrm{x}^2(\mathbf{r},\omega)+\chi_\mathrm{y}^2(\mathbf{r},\omega)+\chi_\mathrm{z}^2(\mathbf{r},\omega)%
\nonumber\\%
\overrightarrow{n}&=&\overrightarrow{n}(\omega)=%
\overrightarrow{\chi }(\mathbf{r},\omega)/\chi (\mathbf{r},\omega ).%
\label{A.22}%
\end{eqnarray}
We additionally assume (and it is a crucial point of the derivation) that the atomic polarization is homogeneous along the atomic sample such that the ``director'' $\overrightarrow{n}(\omega )$ is conservative (does not depend on $\mathbf{r}$) along the path in the phase integrals representation of the Green's function [Eqs.~(\ref{A.15},\ref{A.20})]. That is a relevant approximation in many applications and it generally assumes that the principle directions of the dielectric susceptibility tensor are conserving on a spatial distance scaled by the extinction length $l_{\mathrm{ex}}$.

The asymptotic behavior of the advanced-type photon propagator ${\cal D}_{\mu_1\mu_2}^{(A)}(\mathbf{r}_1,\mathbf{r}_2;\omega)$ is given by complex conjugation and matrix transpose of expression (\ref{A.15}) for the retarded-type propagator. In accordance with the discussion in section \ref{II.A}, the knowledge of the photon propagators gives us the needed mathematical tool for further Monte-Carlo simulations of the scattered wave dynamics in a disordered atomic gas [Eq.~(\ref{2.5})].

\section{Bethe-Salpeter equation in analytical form}\label{Appndx_B}

In accordance with the arguments presented in the main text, see section \ref{Sec.III.B} the pumping process initiated by the spontaneous Raman emission is expressed by the same diagram as amplification term in the imaginary part of self-energy operator for the retarded-type photon propagator. In analytical form the correlation function of the light source is given by the following convolution/matrix transform
\begin{eqnarray}
D^{(S)}_{11'}(\omega)&\!\!\!\propto&\!\! {\cal D}^{(R)}_{12}(\omega)\!\ast\!\!\left[\chi^{(\mathrm{>})}(\omega)-\chi^{(\mathrm{<})}(\omega)\right]_{22'}%
\!\!\!\ast\!{\cal D}^{(A)}_{2'1'}(\omega),
\nonumber\\%
&&%
\label{B.1}%
\end{eqnarray}
where asterisks denote here and below the integral over spatial (if necessary) and sum over polarization variables. For the sake of simplicity we incorporated all the notation details into the numerical arguments such that $1\equiv \mu_{1},\mathbf{r}_1, \ldots$ etc. The advanced-type photon propagator is expressed by hermitian conjugation of the retarded-type propagator such that ${\cal D}^{(A)}_{2'1}={\cal D}^{(R)\ast}_{12'}$.  The sign of proportionality in transformation (\ref{B.1}) indicates any required factors associated with relevant scaling and geometry.

The internal part in Eq.(\ref{B.1}) can be extracted from the expression for the susceptibility tensor (\ref{A.8}). If the upper vertex in the source diagram of graph equation (\ref{2.5}) advances the lower vertex than we have
\begin{eqnarray}
\lefteqn{\chi_{\mu\mu'}^{(\mathrm{>})}(\mathbf{r},\omega)=%
-\frac{i}{\hbar}\sum_{m,m'}\sum_{n,n'}n_0(\mathbf{r})\left(d_{\mu}\right)_{mn}%
\left(d_{\mu'}\right)_{n'm'}}%
\nonumber\\%
&&\times\sum_{n''}\int_{-\infty}^{\infty}\frac{dE}{2\pi\hbar}{\cal G}_{nn''}^{(R)}(E)\bar{\rho}_{n''n'}{\cal G}_{m'm}^{(A)}(E-\hbar\omega)%
\label{B.2}%
\end{eqnarray}
and in alternative situation when the upper vertex retards the lower one we have
\begin{eqnarray}
\lefteqn{\chi_{\mu\mu'}^{(\mathrm{<})}(\mathbf{r},\omega)=%
\frac{i}{\hbar}\sum_{m,m'}\sum_{n,n'}n_0(\mathbf{r})\left(d_{\mu}\right)_{mn}%
\left(d_{\mu'}\right)_{n'm'}}%
\nonumber\\%
&&\times\sum_{n''}\int_{-\infty}^{\infty}\frac{dE}{2\pi\hbar}\bar{\rho}_{n n''}{\cal G}_{n''n'}^{(A)}(E){\cal G}_{m'm}^{(R)}(E-\hbar\omega).%
\label{B.3}%
\end{eqnarray}
The anti-hermitian part of these terms (considered as operators in atomic subspace) defines the rate of local spontaneous Raman emission at frequency $\omega$. The density matrix components of the upper state for $F=3,2$ accumulate the dynamics of the excitation process of the atoms from the ground state $F_0=2$ and can be calculated as a steady state solution of the relevant kinetic master equation. The mean frequency of the emitted photons coincides with the frequency of the $F_0=3\to F=4$ closed transition but the photons are spectrally distributed in the vicinity of their carrier frequency because of the instability of the atomic energy levels. The spectrum is mainly broadened by the "dressing" of level $F_0=2$ by the control mode, similar to Autler-Townes resonance feature shown in inset of figure \ref{fig3}.

As a zero-order calculation step we keep the above Raman emission distributed inside the atomic sample such that $D^{(S)}_{11'}\equiv D^{(0)}_{11'}$. We indicate it as zero-order in the multiple scattering series associated with the trapping mechanism. Then in the first order correction we have to add the light undergone a single scattering event and the kernel block of equation (\ref{2.5}) generates the following term
\begin{eqnarray}
\lefteqn{D^{(1)}_{11'}(\omega)\propto \sum_{\begin{array}{c} \scriptstyle m_2=m'_2 \\ \scriptstyle \in F_0=2,3 \end{array}}%
\sum_{\begin{array}{c} \scriptstyle m_3,m'_3 \\ \scriptstyle \in F_0=3 \end{array}} \bar{\rho}_{m_3,m'_3}}%
\nonumber\\%
&&{\cal D}^{(R)}_{12}(\omega)\!\ast\!\alpha_{23}(\omega)\!\ast\! D^{(0)}_{33'}(\omega)\!\ast\! \alpha^{\dagger}_{3'2'}\!\ast\! {\cal D}^{(A)}_{2'1'}(\omega).%
\label{B.4}
\end{eqnarray}
In dependence down which final state atom decays either $F_0=3$ or $F_0=2$ this correlation function contributes into either the elastic Rayleigh trapping channel or inverse anti-Stokes scattering channel. The elastic singly scattered part then can be substituted in the kernel block again and generate the secondary scattered wave etc. This iteration process is rapidly converging for the number of scattering orders comparable with optical depth of the sample.

The crucial assumption, which we made, requires that the background state $F_0=3$ is considered in our calculation scheme as approximately stable such that all the multiply scattered terms preserve and transfer the original frequency generated by the source term. The steady state components of the density matrix $\bar{\rho}_{m_3,m'_3}$ are driven by the control and microwave modes and used as external parameters in this part of our calculations. The scattering tensors $\alpha_{23}$ and its hermitian conjugated counterpart $\alpha^{\dagger}_{3'2'}$ in Eq.(\ref{A.21}) belong to the same scatterer and we neglect any possible interference paths. Light amplification is presented by the spontaneous Raman emission contributed into retarded and advanced photon propagators as explained in appendix \ref{Appndx_A}. For positive losses $l_{\mathrm{ls}}^{-1}(\omega)>0$, see definition (\ref{3.4}), the iterative series is always converging, which is demonstrated by our numerical simulations.

\section{Dynamics of the atomic subsystem}\label{Appndx_C}

The single particle density matrix $\hat{\rho}=\hat{\rho}(t)$ describing the dynamics of an arbitrary atom driven by external fields obeys the following Lindblad-type master equation written in the operator form
\begin{eqnarray}
\dot{\hat{\rho}}(t)&=&-\frac{i}{\hbar}\left[\hat{H}_0,\hat{\rho}(t)\right]-\frac{i}{\hbar}\left[\hat{V}(t),\hat{\rho}(t)\right]%
\nonumber\\%
&&-\frac{i}{\hbar}\left[\hat{U}(t),\hat{\rho}(t)\right]+\hat{{\cal R}}\hat{\rho}(t).
\label{C.1}%
\end{eqnarray}
Here the first term is responsible for free dynamics of the atom driven by its internal Hamiltonian $\hat{H}_0$ and the second and third terms describe the interactions with external fields. The interaction operators are given in the rotating wave approximation by
\begin{eqnarray}
\hat{V}(t)&=&-\hat{\mathbf{d}}^{(-)}\mathbf{E}_0\,\mathrm{e}^{-i\omega_c t+i\mathbf{k}_c\mathbf{r}}-\hat{\mathbf{d}}^{(+)}\mathbf{E}_0^*\,\mathrm{e}^{i\omega_c t-i\mathbf{k}_c\mathbf{r}},%
\nonumber\\%
\hat{U}(t)&=&-\hat{\mathbf{m}}^{(-)}\mathbf{H}_0\,\mathrm{e}^{-i\Omega t}-\hat{\mathbf{m}}^{(+)}\mathbf{H}_0^*\,\mathrm{e}^{i\Omega t},%
\label{C.2}
\end{eqnarray}
where $\hat{\mathbf{d}}^{(\mp)}$ and $\hat{\mathbf{m}}^{(\mp)}$ are the rising/lowering components of the electric dipole and magnetic dipole operators; $\mathbf{E}_0$ and $\mathbf{H}_0$ are the complex amplitudes of the electric and magnetic fields respectively. In the case of the microwave field we can ignore its spatial profile since its wavelength is typically much longer than the size of the atomic cloud. In the case of linear polarizations for both fields in direction along the $Z$-axis the interaction matrix elements are defined as follows,
\begin{eqnarray}
V_{n\tilde{m}}&=&\left(d_{z}\right)_{n\tilde{m}}E_0,%
\nonumber\\%
U_{m\tilde{m}}&=&\left(m_{z}\right)_{m\tilde{m}}H_0,%
\label{C.3}%
\end{eqnarray}
where we follow our convention and specify by $n,n',\ldots$ any upper states and by $m,m',\ldots$ any ground states. If, as in matrix elements (\ref{C.3}), we need to distinguish the ground states belonging to different hyperfine sublevels we additionally overscribe by tilde the states belonging to the lower sublevel $F_0=2$.

The spontaneous radiative decay of the upper states and of the optical coherences contributes in Eq.(\ref{C.1}) by the following relaxation terms
\begin{eqnarray}
\left(\hat{{\cal R}}\hat{\rho}(t)\right)_{nn'}&=&-\gamma\, \hat{\rho}_{nn'}(t),%
\nonumber\\%
\left(\hat{{\cal R}}\hat{\rho}(t)\right)_{nm}&=&-\frac{\gamma}{2}\, \hat{\rho}_{nm}(t),%
\label{C.4}%
\end{eqnarray}
where $\gamma$ is the natural radiative decay rate. The optical pumping repopulation process providing the atomic polarization transfer from the upper to the ground state via spontaneous decay is described by the following income-type term
\begin{eqnarray}
\lefteqn{\left(\hat{{\cal R}}\hat{\rho}(t)\right)_{m'm}=\gamma\,\sum_{n'n}\rho_{n'n}(t)\sum_{q}C_{F'_0M'_0 1q}^{F'M'}C_{F_0M_0 1q}^{FM}}%
\nonumber\\%
&&\times (-)^{F_0-F'_0}\left[(2F'_0+1)(2F_0+1)\right]^{1/2}(2J+1)%
\nonumber\\ \nonumber\\%
&&\times\left\{\begin{array}{ccc} S & I & F'_0\\ F' & 1 & J \end{array}\right\}%
\left\{\begin{array}{ccc} S & I & F_0\\ F & 1 & J \end{array}\right\}%
\label{C.5}%
\end{eqnarray}
where $m=F_0,M_0$, $m'=F'_0,M'_0$ and $n=F,M$, $n'=F',M'$. Other quantum numbers specifying the atomic states in (\ref{C.5}) are $S=1/2$, $I=5/2$ and $J=3/2$, the electronic spin, nuclear spin and total electronic angular momentum of the upper state respectively. We follow the definitions of the Clebsch-Gordan coefficients and 6j-symbols from \cite{VMK}.

If the excitation of the atomic subsystem is weak such that the initial equally populated Zeeman states for the hyperfine sublevel $F_0=3$ are only slightly perturbed by the external fields, then we can make an analytical estimate for the first order deviation of the steady state density matrix from its initial distribution $\rho^{(0)}_{mm}=(2F_0+1)^{-1}$. But in the general case the system (\ref{C.1}) can be solved only numerically and the steady state components of the density matrix can be further used as external parameters in evaluation of the susceptibility tensor (\ref{A.8}).

\end{document}